# Precise 3D track reconstruction algorithm for the ICARUS T600 liquid argon time projection chamber detector.

M. Antonello[a], B. Baibussinov[b], P. Benetti[c], E. Calligarich[c], N. Canci[a], S. Centro[b], A. Cesana[e],
K. Cieslik[f], D. B. Cline[g], A. G. Cocco[d], A. Dabrowska[f], D. Dequal[b], A. Dermenev[h], R. Dolfini[c],
C. Farnese[b], A. Fava[b], A. Ferrari[i], G. Fiorillo[d], D. Gibin[b], S. Gninenko[h], A. Guglielmi[b],
M. Haranczyk[f], J. Holeczek[j], A. Ivashkin[h], J. Kisiel[j], I. Kochanek[j], J. Lagoda[k], S. Mania[j],
A. Menegolli[c], G. Meng[b], C. Montanari[c], S. Otwinowski[g], A. Piazzoli[c], P. Picchi[l], F. Pietropaolo[b],
P. Plonski[m], A. Rappoldi[c], G.L. Raselli[c], M. Rossella[c], C. Rubbia[a,i], P. Sala[e], A. Scaramelli[e],
E. Segreto[a], F. Sergiampietri[n], D. Stefan[a,*], J. Stepaniak[k], R. Sulej[k,a,*], M. Szarska[f], M. Terrani[e],
F. Varanini[b], S. Ventura[b], C. Vignoli[a], H. Wang[g], X. Yang[g], A. Zalewska[f], K. Zaremba[m].

**(ICARUS Collaboration)**

a   Laboratori Nazionali del Gran Sasso dell'INFN, Assergi (AQ), Italy
b   Dipartimento di Fisica e Astronomia e INFN, Università di Padova, Via Marzolo 8, I-35131 Padova, Italy
c   Dipartimento di Fisica e INFN, Università di Pavia, Via Bassi 6, I-27100 Pavia, Italy
d   Dipartimento di Scienze Fisiche e INFN, Università Federico II, Napoli, Italy
e   INFN, Sezione di Milano e Politecnico, Via Celoria 16, I-20133 Milano, Italy
f   Henryk Niewodniczanski Institute of Nuclear Physics, Polish Academy of Science, Krakow, Poland
g   Department of Physics and Astronomy, University of California, Los Angeles, USA
h   INR RAS, prospekt 60-letiya Oktyabrya 7a, Moscow 117312, Russia
i   CERN, CH-1211 Geneve 23, Switzerland
j   Institute of Physics, University of Silesia, Uniwersytecka 4, 40-007 Katowice, Poland
k   National Centre for Nuclear Research, A. Soltana 7, 05-400 Otwock/Swierk, Poland
l   Laboratori Nazionali di Frascati (INFN), Via Fermi 40, I-00044 Frascati, Italy
m   Institute of Radioelectronics, Warsaw University of Technology, Nowowiejska 15/19, 00-665 Warsaw, Poland
n   INFN, Sezione di Pisa, Largo B. Pontecorvo 3, I-56127 Pisa, Italy
*   corresponding authors: Dorota.Stefan@lngs.infn.it, Robert.Sulej@lngs.infn.it

**Abstract**

Liquid Argon Time Projection Chamber (LAr TPC) detectors offer charged particle imaging capability with remarkable spatial resolution. Precise event reconstruction procedures are critical in order to fully exploit the potential of this technology. In this paper we present a new, general approach to 3D reconstruction for the LAr TPC with a practical application to the track reconstruction. The efficiency of the method is evaluated on a sample of simulated tracks. We present also the application of the method to the analysis of stopping particle tracks collected during the ICARUS T600 detector operation with the CNGS neutrino beam.

## 1. Introduction.

The LAr TPC detector idea, proposed in 1977 by C. Rubbia [1], provides spatial and calorimetric measurement of charged particles with the level of details comparable to bubble chamber technology. The concept is now exploited in several projects around the world [2, 3, 4, 5] with the ICARUS T600 [6, 7] being the largest presently operating detector, located at Gran Sasso underground National Laboratory, on the CNGS neutrino beam. The LAr TPC technique is of interest for a wide physics program including studies of neutrino oscillation parameters, sterile neutrinos, CP violation, violation of baryonic number conservation and dark matter searches.

The operating principle of the LAr TPC detector as implemented in the ICARUS T600 is illustrated in Fig. 1a. A charged particle produces ionization electrons and scintillation light along its path. Free electrons drift in a uniform electric field toward the anode that is composed of three wire planes. Diffusion of the drifting electrons is low enough, 4.8 cm$^2$/s, [7], to preserve details of the ionizing particle track. A signal is induced in a non-destructive way on the first two wire planes, *Induction1* and *Induction2*, which are practically transparent to the drifting electrons. The signal on the third wire plane, *Collection*, is formed by collecting the ionization charge, which is also the source of the calorimetric measurement. Different orientation of the wires in the anodic planes (0°, +60°, -60° with respect to the horizontal, with 3 mm wire spacing in each plane) allows localization of the signal source in the XZ plane as shown in Fig. 1b, while the Y coordinate, which defines the distance to the wire planes, is calculated from the wire signal timing and the electron drift velocity, 1.59 mm/µs. Wire signals are amplified and digitized with 2.5 MHz sampling frequency which results in 0.64 mm spatial resolution along the drift coordinate. The absolute event timing, $t_0$, is provided by the prompt signal from the photomultipliers collecting the scintillation light. Finally, digitized waveforms from consecutive wires placed next to each other form 2D projection images of an event, like in the example of neutrino interaction shown in Fig. 2.

The reconstruction of an event is split into a series of several steps realized with independent algorithms. It starts with the identification of individual signals on wires, so called hits. At this point the position and calorimetric information are assigned to hits. In the next step hits are aggregated in clusters forming 2D structures: tracks and showers. Hit clustering is a challenging image recognition task in itself given the complexity and variety of possible event topologies. A solution based on the *DBSCAN* algorithm has been proposed in [2], however the techniques that are efficient at reconstructing complex topologies are being under study. Independently from the clustering method, 3D objects are reconstructed using 2D hit clusters associated in at least two wire planes. 3D reconstruction provides information necessary for further analysis, especially directions of particles and their momenta are crucial for the reconstruction of the event kinematics. Also the particle identification of low energy hadrons is based on the energy losses along their tracks, d$E$/d$x$, and it is dependent on the precision of the spatial reconstruction.

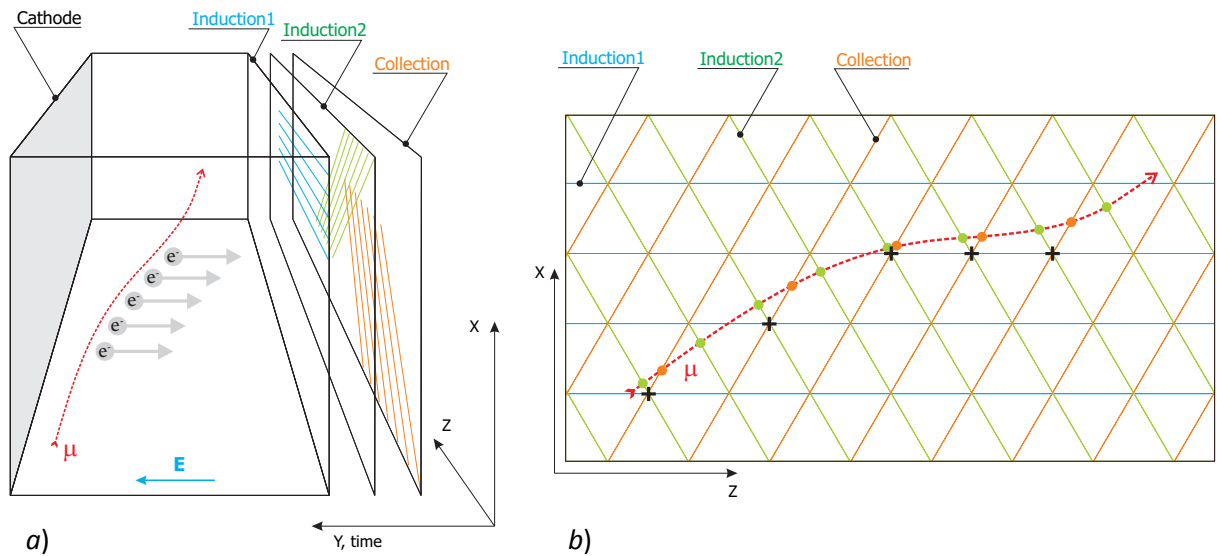

Fig. 1. Schematic view of the ICARUS T600 readout principle; one of four TPC's is shown: *a*) 3D view with marked µ track and the ionization electrons e$^-$ drifting in the electric field E toward the readout wire planes; *b*) the XZ projection with marked actual intersections of the particle track and readout wires (green and orange points) and an example of points on the reconstructed track that may be obtained by associating wire signals from *Induction2* and *Collection* planes using drift timing (black crosses).

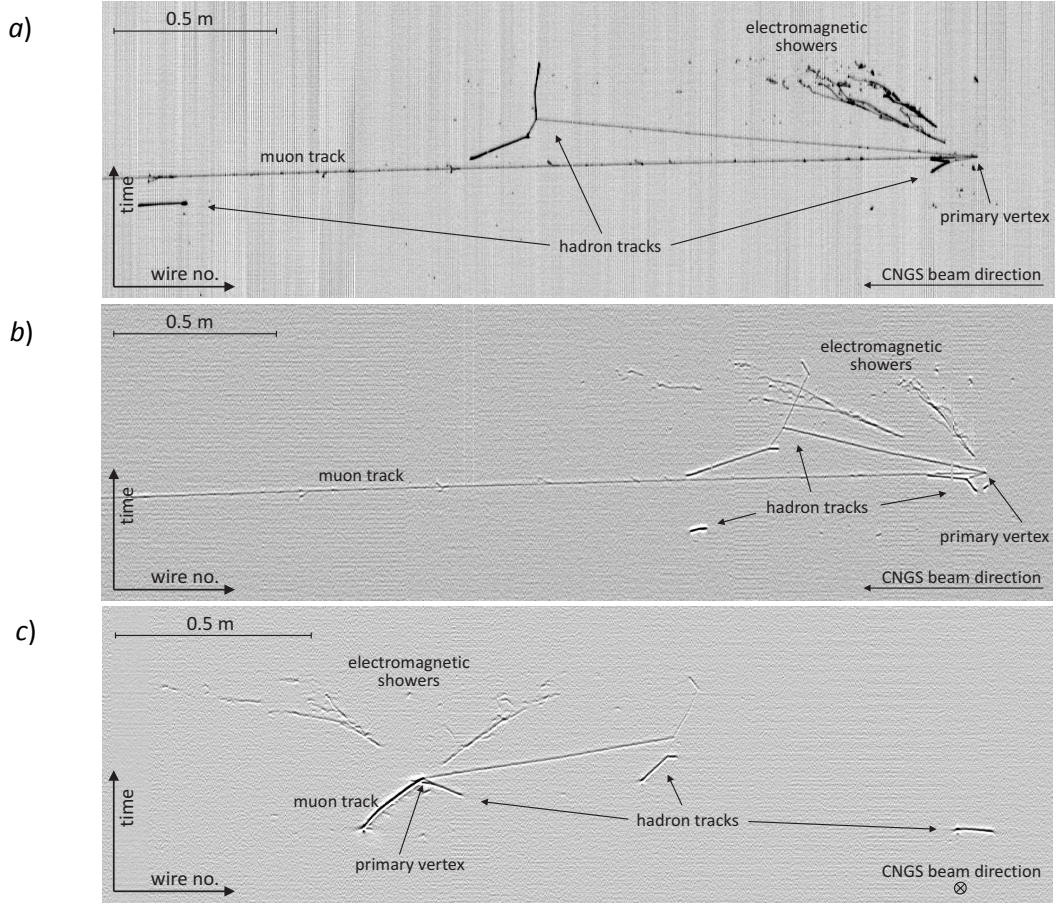

Fig. 2. Data, CNGS $\nu_\mu$ charge current interaction in the ICARUS T600 detector, a fragment of the wire data in one of the detector modules: *a*) *Collection* plane; *b*) *Induction2* plane; *c*) *Induction1* plane. The horizontal wires of the *Induction1* plane form the projection on XY plane and give the frontal view of the event while the *Collection* and *Induction1* planes form YZ projections and give the top views of the event seen at two angles.

Typical approaches to the track 3D reconstruction presented in recent papers [2, 3, 6, 7] are based on matching the hits or the track end points from the individual 2D wire planes by their drift timing. Examples of reconstructed events were shown, however without the evaluation of the algorithm efficiency on a large sample of tracks.

The first limitation of the approach based on the drift time matching is illustrated in Fig. 1b. The XZ position of the 3D point is obtained as the crossing point of the wires containing the matched pair of 2D hits. This leads to a position quantization in the XZ plane, 3.5x3.5 mm$^2$ assuming 3 mm wire to wire spacing and 60° inclination between wires of the consecutive readout planes. Moreover, any spurious hit matching adds an additional error of 3.5 mm per each wire shift with respect to the correct matching. This effect introduces kinks and distortions to the track constructed with matching hit by hit, visible also in the track examples shown in [6, 7]. On the other hand, the straight line approximation is not sufficient to reproduce the particle trajectory in all details available with LAr technology, such as elastic scatterings of the low energy particles.

Another drawback of the approach based on the drift time matching is the inefficiency of the reconstruction of tracks parallel to the wire planes. The low variation in the hit drift timing along the track increases ambiguities in the individual hit association between the wire planes.

The disadvantages that we also observed during the studies of the drift time based matching of 2D projections include:
  a) incomplete information, e.g. missing parts of the track in one of the wire planes due to a hardware problem, is difficult to manage;
  b) simultaneous use of the three wire planes is not straightforward and was not developed according to our knowledge;
  c) a robust algorithm requires pre-processing, such as hit sorting, in each 2D projection independently before the projections matching; we found this difficult in case of the sophisticated track topologies, particularly in case of scatterings at a sharp angle in the 2D projection.

In data, tracks with spurious reconstruction are difficult to distinguish from tracks with correctly reconstructed features. This motivated us to search for a different concept of 3D reconstruction that would be capable of reproducing details along the whole track, at any track orientation.

In this paper we propose to build 3D objects by simultaneous optimization of their 2D projections to match data in the wire planes. A single fit function is constructed to combine all pieces of information available in data with the constraints specific to the considered object type. The proposed new approach is applied in this paper to tracks, but the idea is general and may be explored in the reconstruction of cascade-like objects, detailed reconstruction of interaction vertex regions as well as in global event reconstruction. It is also straightforward to adapt the implementation to detector readout concepts different from wire planes, e.g. based on Large Electron Multipliers (LEM) [8].

In Section 2 we describe the general algorithm idea and its practical implementation details. Then we show (Section 3) the algorithm performance evaluated on simulated particle tracks, compared with the results obtained from the previously developed hit matching algorithm. The proposed method is cross-checked on a sample of stopping particles selected in the events collected during the CNGS run (Section 4). Finally, in Section 5, we summarize results and comment on possible future enhancements.

## 2. Algorithm.

The particle real track $T$ is observed in the detector as a set of three 2D projections $P_{I1}(T)$, $P_{I2}(T)$, $P_C(T)$ to *Induction1*, *Induction2* and *Collection* wire planes respectively. In practice these projections consist of 2D hits. The 3D fit trajectory $F$ may be projected to the wire planes according to the same operators $P(F)$. We propose to build the fit $F$ by minimizing a measure of the distance D between the fit projections and the track hits in all wire planes simultaneously, with constraints $C_j(F)$ that may include factors such as trajectory curvature and distance to the already identified and reconstructed interaction vertices. This may be expressed with an objective function $G(F)$:

$$G(F) = \sum_i \alpha_i \, \mathrm{D}[\mathrm{P}_i(T), \mathrm{P}_i(F)] + \sum_j \beta_j \, \mathrm{C}_j(F), \qquad (1)$$

where wire planes, denoted with the index *i*, and constraint factors, denoted with the index *j*, have a weighted impact on the overall G(F) value according to the $\alpha_i$ and $\beta_j$ coefficients. For the practical implementation of constructing the best fit track we have adopted the *Polygonal Line Algorithm* [9], PLA, described in more details in Section 2.2, however any other approach, even straight line approximations may follow the same concept of simultaneous optimization of 2D projections.

2.1. Hit reconstruction.

We briefly describe the method of the individual hit reconstruction since it is the essential base for the spatial and calorimetric measurements.

In the first stage hits are identified in all wire planes following the algorithm presented in [7]. As a result hit positions in the drift coordinate are obtained together with the drift time windows, which cover the hit ADC waveform with a margin required for the second stage.

In the second stage fitting hits within the drift time windows is applied to the *Collection* plane. This improves the hit positioning, resolves overlapping hits and allows the reconstruction of the individual hit energy deposits.

Hits with adjacent drift time windows form a group that is fitted as it would be for a single window. The fit function for each drift time window reads

$$f_w(t) = B + \sum_{i=1}^{N_p} p_i(t), \quad p_i(t) = A_i\, e^{-(t-tp_i)/\tau_1} \left(1 + e^{-(t-tp_i)/\tau_2}\right)^{-1}, \qquad (2)$$

where: p(*t*) is the function describing the impulse response of the wire readout electronics, later in text called a pulse; $A_i$ is the $i^{th}$ pulse amplitude; $tp_i$ is the $i^{th}$ pulse timing; $\tau_1$ and $\tau_2$ are the rising and falling time constants respectively with common values for all pulses in the fit; *B* is the fit baseline. The fit parameter values and the number of pulses, $N_p$, are iteratively optimized to minimize $\chi^2 = \Sigma_t [ADC(t) - f_w(t)]^2/N_{free}$ where $N_{free} = N_{ADC} - (2N_p + 3)$ is the number of degrees of freedom of the fit, $N_{ADC}$ is the number of ADC samples in the drift time window, $2N_p+3$ is the total number of $f_w(t)$ parameters. Pulses $p_i(t)$ become the new hits that replace those initially found at the hit identification stage. The maxima of pulses are considered as the drift coordinate positions of the new hits, $t_d$.

The particle energy deposit observed in a hit is calculated as $q_E = q_C W\, e^{t_d/\tau_e}$ eV, where $q_C$ is the hit charge calculated as the integral of the corresponding pulse p(*t*) scaled according to the calibration factor $152 \pm 2 \times 10^{-4}$ fC/(ADC×μs), as evaluated in [10]. *W* is the average energy needed for the creation of an electron-ion pair, $23.6^{+0.5}_{-0.3}$ eV [11], and $\tau_e$ is the free electron lifetime, which is monitored during the detector operation, as presented in [6].

Particle tracks perpendicular to the drift direction produce signals that are fitted with a single pulse per wire, while wire signals induced by tracks more parallel to the drift direction need to be considered as sequences of pulses with varying amplitudes and timings (Fig. 3).

The uncertainty on the energy deposit of the reconstructed hits was estimated from reconstruction of the simulated signal with added electronics noise based on data. The uncertainty is almost independent from the actual hit energy deposit value and it is constant at the level of $\sigma = 0.06$ MeV (Fig. 4). The lowest amplitude hits observed on the particle tracks are related to the minimum ionizing muons; energy deposit of such hits is on the level of 0.5-0.7 MeV. The resulting worst case inaccuracy of the energy deposit per hit is roughly 10%.

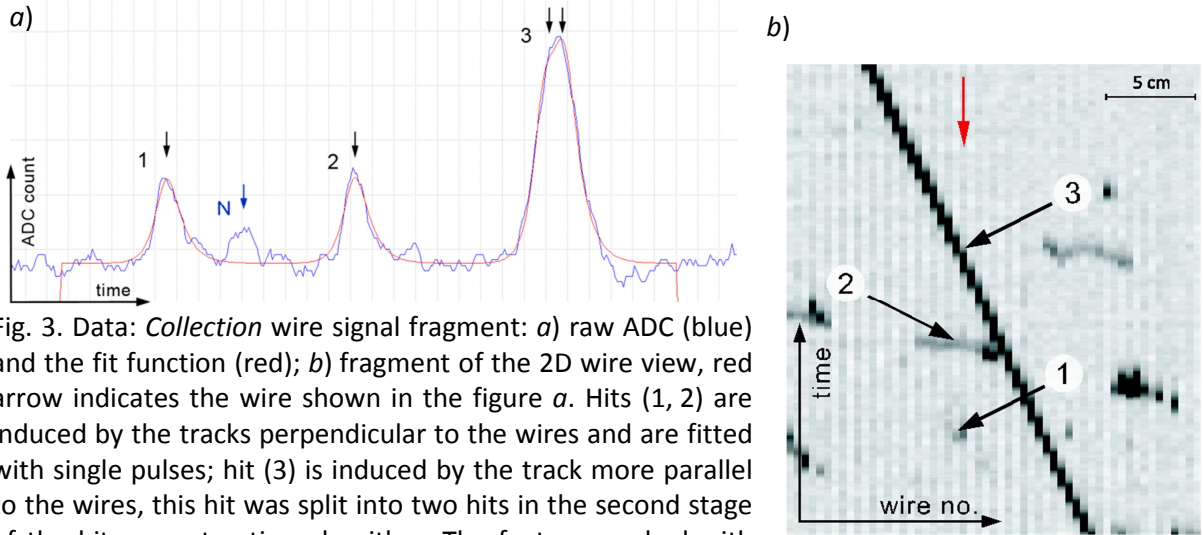

Fig. 3. Data: *Collection* wire signal fragment: *a)* raw ADC (blue) and the fit function (red); *b)* fragment of the 2D wire view, red arrow indicates the wire shown in the figure *a*. Hits (1, 2) are induced by the tracks perpendicular to the wires and are fitted with single pulses; hit (3) is induced by the track more parallel to the wires, this hit was split into two hits in the second stage of the hit reconstruction algorithm. The feature marked with (N) in the raw ADC is too low to be recognized as hit during the signal fit optimization.

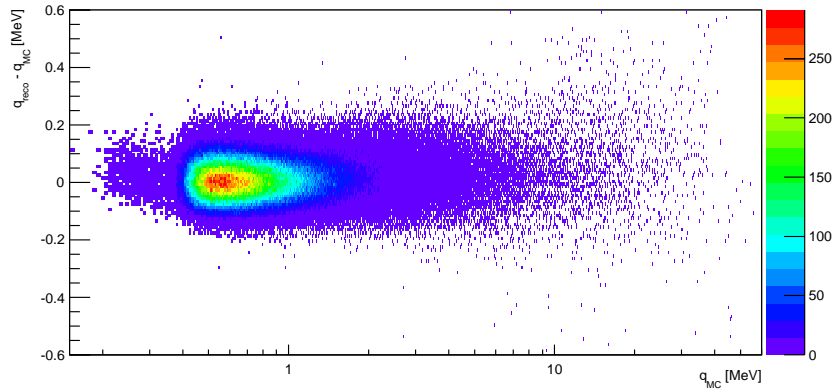

Fig. 4. Simulation: the absolute difference between the reconstructed hit energy deposit ($q_{\text{reco}}$) and the simulated hit energy deposit ($q_{\text{MC}}$) as a function of $q_{\text{MC}}$. The hits shown are originating from muon tracks simulated with isotropic initial direction. The tail of the distribution with $q_{\text{MC}} < 0.4$ MeV corresponds to the hits at track end-points, only partially crossing the distance between consecutive wires. The peak of the distribution corresponds to the hits of minimum ionizing particles; the points at high values of $q_{\text{MC}}$ come from low energy, highly ionizing parts of tracks.

2.2. Track reconstruction.

The objective function $G(F)$ (Eq. 1) has been inspired by the PLA formulation, which is an efficient algorithm for the principal curve finding problem [9]. The principal curve, in our case the best track fit $F$, is approximated with the polygonal line determined by 3D points, called later nodes, interconnected with straight 3D segments. The difference from the original PLA application is that we are looking for the principal curve in 3D while the distance to data points is given in 2D projections. The track fit shape and its projections distances to the 2D hits need to be optimized to allow unbiased reconstruction of the track features, such as scatterings and decay points, taking into account also the hit positions accuracy.

The diagram in Fig. 5 shows the main steps of the PLA algorithm. In the initialization the first segment with two nodes is placed in the 3D space, according to the procedure described in Section 2.2.1. Then the track fit $F$ is constructed in an iterative way by adding new nodes and rebuilding segments. After adding a new node, the positions of all nodes, denoted with $\mathbf{n}_k$ where $k = 1…K$, are optimized in the inner loop of the algorithm. Simultaneously with changes of the node positions, the fit 2D projections to the wire planes are updated.

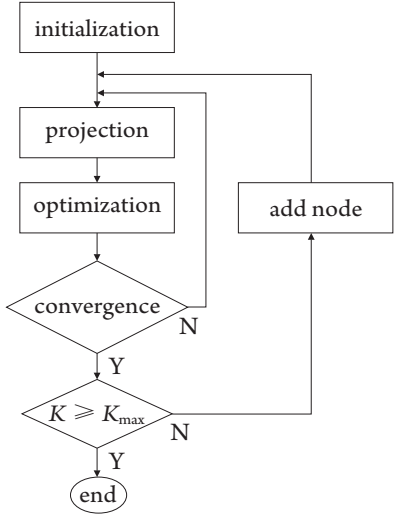 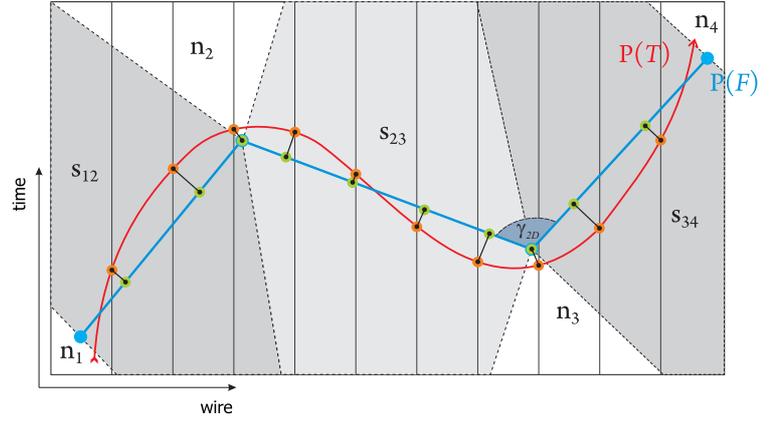

Fig. 5. Fit optimization diagram. The inner loop contains the hit assignment to nodes and segments of the fit (*projection* step) and the optimization of node 3D positions. New nodes are introduced to the fit in the outer loop until $K_{max}$ number of nodes is reached.

Fig. 6. Schematic view of the track P($T$) (red) and fit P($F$) (blue) projections in the wire plane. 2D hits (orange dots) are marked on the track projection intersections with wires; hit projections to P($F$) are marked as green dots; hit distances to P($F$) are marked with black sections. Hits from grey shaded regions $s_{xy}$ are assigned to segments; hits from white regions $n_x$ are assigned to nodes. Indicated angle $\gamma_{2D}$ is the 2D projection of the $\gamma$ angle between the fit segments in 3D (see Eq. 7 in text).

The 2D projections of the track fit $F$ are determined by the node 2D projections, $P_i(\mathbf{n}_k)$, which also describe segment 2D projections. The algorithm stops when the maximum number of nodes, $K_{max}$, is reached.

The inner loop of the algorithm consists of two steps: the projection and the optimization. In the projection step, 2D hits are assigned to the fit segment or node. This is done by finding the segment/node 2D projections with the minimal distance to the 2D hit, as it is illustrated in Fig. 6.

During the optimization step the node positions are updated to minimize the objective function (Eq. 1). To reduce the computational complexity we use a local version of the objective function[1]: $G(F) = \Sigma_k g(\mathbf{n}_k)$ where $g(\mathbf{n}_k)$ is a function assigned to the $k^{th}$ node of the fit:

$$g(\mathbf{n}_k) = d(\mathbf{n}_k) + \beta_v c_v(\mathbf{n}_k) + \beta_a c_a(\mathbf{n}_k) \qquad (3)$$

and it consists of the following components:

a) weighted average of the squared distance of the fit projection to the 2D hits that are assigned to the $k^{th}$ node and segments connected to that node, in all wire planes:

---

[1] We use a gradient descent method with the finite differences approximation of the gradient $\partial G(F)/\partial \mathbf{n}_k$, therefore it is convenient to express the objective function as a sum of independent components.

$$\mathrm{d}(\mathbf{n}_k) = \frac{1}{N_k} \sum_i \left( \alpha_i \sum_{n \in N_{ik}} \omega_n \left\| \mathbf{h}_n^i - \mathrm{f}_{2\mathrm{D}}(\mathbf{h}_n^i) \right\|^2 \right), \qquad (4)$$

where $N_k$ is the number of considered hits, $N_{ik}$ includes indices of hits in the $i^{th}$ wire plane among all hits considered with $k^{th}$ node; $\mathbf{h}_n^i$ is the position of the $n^{th}$ hit in the wire plane $i$ and $\mathrm{f}_{2\mathrm{D}}(\mathbf{h}_n^i)$ is the position of this hit projection to the fit 2D projection, $P_i(F)$, as illustrated in Fig. 6; $\omega_n$ is the coefficient used to weight the contribution of the individual hits to $\mathrm{d}(\mathbf{n}_k)$ value; we calculate $\omega_n$ as the ratio of $n^{th}$ hit amplitude over the maximum hit amplitude among hits considered in the inner sum of (Eq. 4), which efficiently supresses the impact of the noise hits spuriously accounted to the track, usually of the much lower amplitudes than the signal hits;

b) the average squared distance of the fit to the 3D vertices created independently from the track reconstruction algorithm:

$$\mathrm{c}_v(\mathbf{n}_k) = \frac{1}{M} \sum_{m \in M_k} \left\| \mathbf{v}_m - \mathrm{f}_{3\mathrm{D}}(\mathbf{v}_m) \right\|^2, \qquad (5)$$

where $M$ is the number of all used vertices, $M_k$ includes indices of vertices assigned to the $k^{th}$ node and segments connected to that node, $\mathbf{v}_m$ is the position of the $m^{th}$ vertex and $\mathrm{f}_{3\mathrm{D}}(\mathbf{v}_m)$ is the position of this vertex projection to the fit; in the present stage we do not differentiate accuracy of the individual vertex positions, however it is straightforward to introduce a coefficient similar to $\omega_n$ in (Eq. 4) when the reconstruction accuracy is available; identification and reconstruction of the 3D vertices is beyond the scope of this paper, therefore we only note that such vertices may tag the particle interaction, decay points or delta ray spots identified along the particle track;

c) constraint on the angles $\gamma$ between the consecutive 3D segments, and on the length of the outermost segments where the angle $\gamma$ cannot be calculated:

$$\mathrm{c}_a(\mathbf{n}_k) = K^p \left[ \pi(\mathbf{n}_{x=k-1}) + \pi(\mathbf{n}_{x=k}) + \pi(\mathbf{n}_{x=k+1}) \right], \qquad (6)$$

$$\pi(\mathbf{n}_x) = \begin{cases} r_x^2 (1 + \cos \gamma_x) & x \in \langle 2; K-1 \rangle & \text{inner nodes} \\ \eta \cdot \left\| \mathbf{n}_{x+1} - \mathbf{n}_x \right\|^2 & x = 1 & \text{first node} \\ \eta \cdot \left\| \mathbf{n}_x - \mathbf{n}_{x-1} \right\|^2 & x = K & \text{last node} \\ 0 & x < 1 \text{ or } x > K \end{cases} \qquad (7)$$

where $r_x$ is the radius of the hits assigned to the $x^{th}$ node and connected segments (the maximum distance between the hit and the mean position of all considered hits), $\eta$ is an arbitrary factor assigned to the end nodes[2]; scale factor $p$ allows to shape the contribution of the constraint while $K$ (the number of the nodes) is growing[3]; the constraint on the fit segment angles allows to keep a smooth track fit along the particle trajectory.

Coefficients $\alpha_i$, $\beta_a$, $\beta_v$ in (Eq. 3, 4) correspond to $\alpha_i$ and $\beta_i$ in (Eq. 1) and allow to keep balance between over-fitting to the noise in hit/vertex positions and ability to reconstruct correctly the significant track features. The actual values of these coefficients depend on the noise conditions of the readout wire planes[4]. The performance in the localization of track features, scattering and decay points, is illustrated in the examples of simulated tracks in Section 3 and in the example of decaying kaon in Section 4.

---

[2] $\eta$ value was determined empirically as 0.05 to avoid excessive stretching of the end segments or shortening the whole track fit (however precise tuning has no strong effect on the obtained fit).
[3] We use $p = 1.8$ to allow slow decrease of the constraint while approaching the minimum of the objective function.

The node positions optimization step is finished when the minimization of G($F$) converges to a stable value[5]. Then the new node is added to the segment with the maximum number of hits assigned in the projection step. The position of the new node is chosen as the one that separates the selected segment into two parts containing the same number of hits.

The stopping condition of the algorithm, the maximum number of nodes $K_{max} = \min(N/5, 7N^{1/3})$, is based on the number of hits $N$; it reflects the track length and allows to keep an higher number of hits per segment for high energy particles with long tracks. Short tracks with $N \leq 5$ are approximated with a single segment. Stopping condition parameters were optimized to maximize the spatial reconstruction efficiency measures shown in Section 3.

Finally, the 3D positions corresponding to the 2D hits are calculated:
a) if the hit is assigned to the fit segment: the hit projection to the fit segment projection, $f_{2D}(\mathbf{h})$, determines the 3D hit position since the relative distance from the segment beginning with respect to the segment length is the same in 2D projection and in 3D space;
b) if the hit is assigned to the fit node: the hit 3D position is simply the node position.

### 2.2.1 Initialization.

The fit optimization starts from two nodes connected with a single segment. The initial 3D positions of the first two nodes should roughly correspond to the track end points. We assume that hits in the individual wire views are not ordered and the exact matching of hits corresponding to the actual track end points is not possible. The initial node positions are evaluated as follows: a straight line is fitted to the hits within the wire plane using the linear regression; two outermost projections of hits to the fitted line are considered as 2D end points; end points from two wire planes are paired by the minimal drift time difference to obtain 3D node positions[6]. Then the minimization of the objective function G($F$) is performed to find the optimal positions of the first nodes. In case of short tracks parallel to wire planes it is possible to obtain wrong matching of 2D end points due to the small drift time difference of two combinations of 2D end point pairs. Then the two possibilities are tested and the one with smaller resulting G($F$) is chosen.

### 2.2.2 Wire plane spacing.

The timing difference between the corresponding signals observed on successive wire planes is the combination of the electron drift time between the wire planes and the delays in the electronics. Inaccurate estimation of the overall delay may cause strong systematic effects in terms of the expected precision of the reconstruction. The delay value for the ICARUS T600 data has been fine-tuned empirically, by reconstructing the tracks using two wire planes and minimizing the distance between fit projection and track hits in the third wire plane (Fig. 7).

---

[4] The coefficient values used for the ICARUS T600 data are the following: $\alpha_{Coll}$ = 1.0, $\alpha_{Ind2}$ = 0.8, $\alpha_{Ind1}$ = 0.2 (for *Collection*, *Induction2* and *Induction1* planes respectively); $\beta_v$ = 1.0; $\beta_a$ = 2.0. Values were adjusted empirically to maximize the reconstruction efficiency.

[5] The relative change of G($F$) is calculated after each step of the minimization algorithm, which updates all node positions. The value of G($F$) is considered as stable when the relative change is below $10^{-4}$, however in the first stages of building the track we use higher values to speed up the computations.

[6] This rough approximation of 3D position may fall out of the actual detector volume. In such a case we simply limit the obtained position to the nearest one inside the detector.

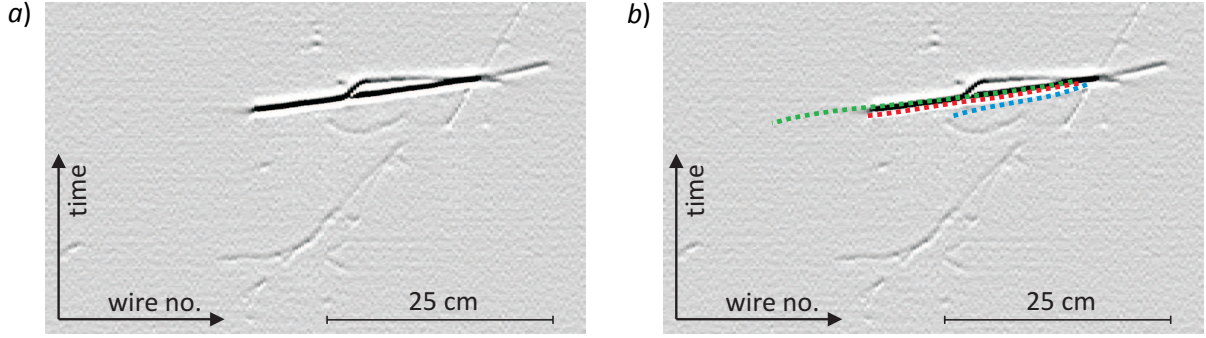

Fig. 7. Muon track (data, CNGS ν_μ CC interaction) observed in the *Induction1* view, and reconstructed using the *Collection* and the *Induction2* plane hits with different delay values between these planes: *a)* *Induction1* raw data view; *b)* 3D track reconstructed with three different delay values and then projected to *Induction1* view, orange – optimal delay, blue and green – over and underestimated delay respectively.

Long tracks, almost parallel to the wire planes, were used in the tuning since they are the most sensitive to changes in the timing delay between planes.

2.2.3 Calorimetric measurement.

The sequence of the ionization charge $dQ$ collected per track $dx$ length, $dQ/dx$, is evaluated from the charges and the 3D positions of the *Collection* plane hits. Since tracks nearly parallel to the drift direction can contain several hits within a small range along the drift direction, it is convenient to limit the minimal $dx$ length to a value close to the wire spacing distance, in our application 2.7 mm. The hit charge is assigned to the $\delta x$ length surrounding the hit, calculated as $(x_1 + x_2)/2$, where $x_1$ and $x_2$ are the distances to the preceding and subsequent *Collection* hit respectively. Hit charges and $\delta x$ lengths are summed up until the minimum value of $dx$ is reached. In this way $dx$ values are comparable for any track orientation with respect to the *Collection* wires direction. Then the correction due to the recombination effect [12] may be applied to obtain the actual value of the energy deposit per track $dx$ length, $dE/dx$, according to the Birk's semi-empirical formula, which can be expressed as:

$$dE = dQ/R, \quad R = A - \frac{k}{\varepsilon \cdot \rho} \cdot \frac{dQ}{dx}, \qquad (8)$$

where R is the correction factor, calculated with the parameters: $A = 0.81$, $k = 0.055$ (kV/cm)(g/cm$^2$)/MeV, $\varepsilon = 0.5$ kV/cm and $\rho = 1.4$ g/cm$^3$ that were obtained from data collected during the detector test run [12].

**3. Algorithm performance on simulated tracks.**

The algorithm performance was evaluated on samples of simulated stopping protons and muons, two species with different track properties: low and medium energy protons produce relatively straight tracks characterized by high ionization while muons of the comparable range are more scattered and give signals of much lower amplitude. Particles were transported in LAr using the FLUKA simulation package [13, 14], including all physical interactions such as delta rays, inelastic scatterings, decays and absorptions. The ionization charge along the track was subject to the recombination effect [12]. In order to evaluate the geometrical reconstruction efficiency the volume of the detector was divided into elements of 3 mm$^3$, called later MC cells, used to accumulate the resulting charge without any instrumental effect. Simultaneously, the charge was also collected in bi-dimensional structures that reproduce the 2D wire views keeping the same wire spacing and orientation as in the real detector and a drift coordinate granularity finer than the one corresponding to the ADC sampling. The resulting charge deposition was convoluted with the readout channel response using packages dedicated to the ICARUS T600 detector. Finally, the electronics noise has been applied to the wire signals, with use of the noise parameters obtained from data. Then the hit and track reconstruction procedures were applied, and compared with the unperturbed MC cells.

The presented tests are the reconstruction of isolated tracks, without added information about vertices. The simple topologies of the simulated tracks allow to show basic properties of the proposed method. The tests include: reconstruction of the particle initial direction based on the comparison with the simulated initial momentum vector, the spatial reconstruction along the track based on MC cells positions and calorimetric reconstruction of the particle kinetic energy. Two examples are shown to illustrate the algorithm properties. The reconstruction results obtained with the proposed method are compared to the results of a previously developed algorithm. This previous approach was based on the independent reconstruction of the two 2D projections of the track, with the application of the PLA [9]. Hits from the track projection in the *Collection* plane were paired with the hits on the track projection in the *Induction* plane using drift timing. Hits ordering obtained from PLA was used to resolve ambiguities in choosing the best matching pair if several hits were found with the close drift time values. Matched hit pairs were used to obtain the 3D track points. The comparison of both methods is shown on the proton track samples in the following subsections.

The effects pointed out in the Introduction, such as the spurious 2D hit matching and the quantization of 3D positions in the XZ plane, are well visible in the reconstruction of tracks nearly parallel to the XZ plane, as it is shown in the example in Fig. 8. These problems are practically eliminated in the proposed algorithm. The second example, shown in Fig. 9, illustrates a narrow angle between two tracks in a decay chain. Such an event topology was not manageable with our implementation of conventional hit matching approach.

3.1  Reconstruction efficiency dependence on the track inclination.

Since the reconstruction depends on the track orientation with respect to the readout wires we present an efficiency evaluation as a function of the angle $\theta_w$ between the normal to the wires direction and the initial particle direction in 2D projections. Samples of 5000 protons and muons were simulated with isotropic direction, with initial kinetic energy $E_{k\,p}$ = 232 MeV and $E_{k\,\mu}$ = 100 MeV respectively, in order to ensure comparable range (30 cm) of both kinds of particles. The particles that are stopping or decaying at rest were selected for further tests.

Cases with all distances of MC cells to the fit below 5 mm were considered as correctly reconstructed tracks. The fraction of correctly reconstructed tracks as a function of $\theta_w$ is shown in Fig. 10. The comparison of results obtained with the new method and the hit matching based approach is shown on the proton track sample in Fig. 10a. The spatial reconstruction is better with the new 3D approach even if applied to the relatively simple proton tracks. The problems pointed in the Introduction and illustrated with the example in Fig. 8 are cause of the significant inefficiencies of the traditional method in the detailed spatial reconstruction of tracks.

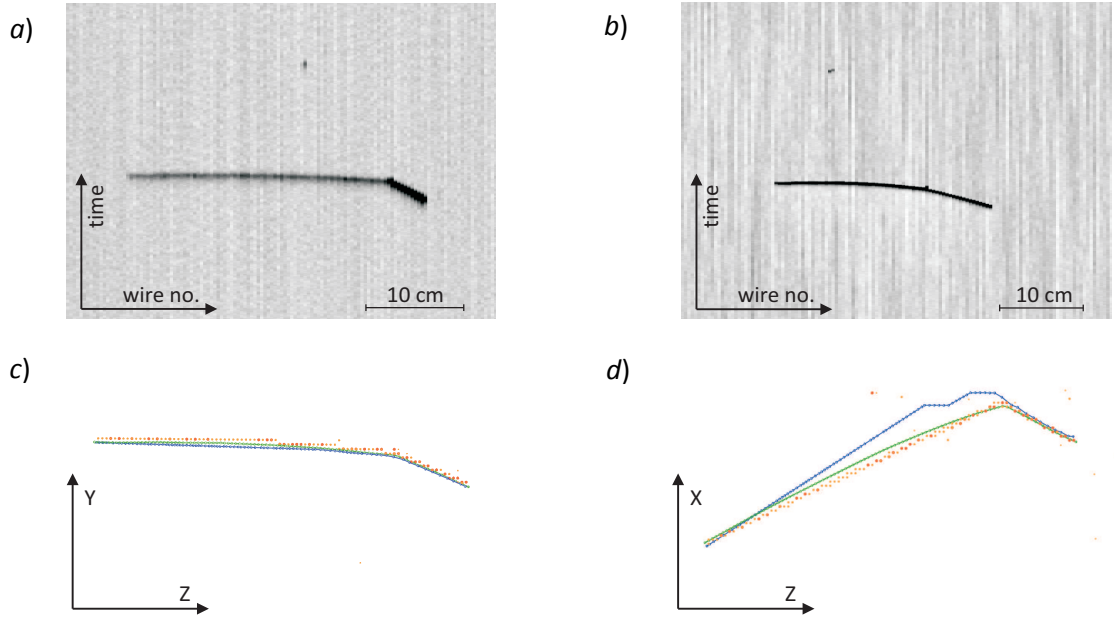

Fig. 8. Simulation and reconstruction of a proton track: *a*) *Collection* wire view; *b*) *Induction2* wire view; *c*) and *d*) respectively YZ and XZ views of the MC cells (orange dots), and the track reconstructed: using matching 2D projections by drift timing (blue curve) and using 3D fit optimized to 2D projections (green curve).

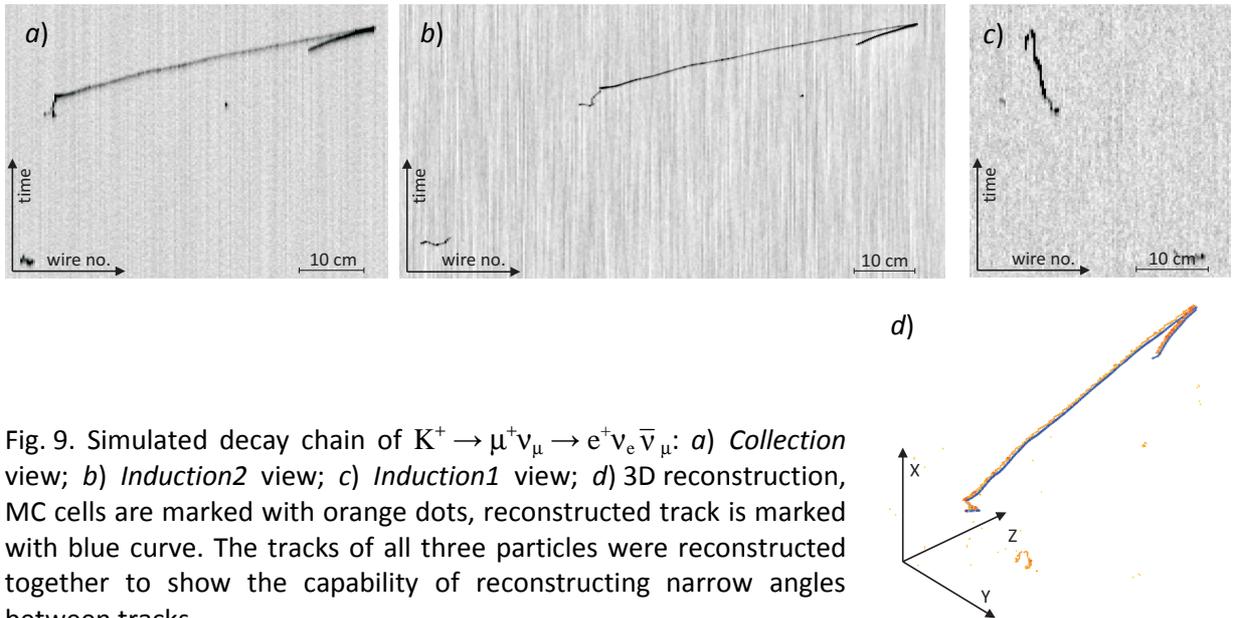

Fig. 9. Simulated decay chain of $K^+ \to \mu^+ \nu_\mu \to e^+ \nu_e \bar{\nu}_\mu$: *a*) *Collection* view; *b*) *Induction2* view; *c*) *Induction1* view; *d*) 3D reconstruction, MC cells are marked with orange dots, reconstructed track is marked with blue curve. The tracks of all three particles were reconstructed together to show the capability of reconstructing narrow angles between tracks.

The worsening of the reconstruction efficiency of the proposed method, seen at $\theta_w = 0°$, 180°, is related to tracks parallel to the wire planes with a short projection (a few hits) in one of the planes. A second source of the inefficiency is the ambiguity of the fit initialization when the straight track in the XZ plane has two possible 3D fit solutions. The inefficiency seen around $\theta_w = 90°$, is related to tracks nearly parallel to the drift direction, which produce long signals on a few wires, more difficult to resolve in the *Induction2* plane. The efficiency obtained for muon tracks is smaller than the one for protons due to the spurious assignment of the muon decay product hits to the muon track and vice versa. The effect caused by the wrong hit assignment is increased at the muon track inclinations close to $\theta_w = 90°$.

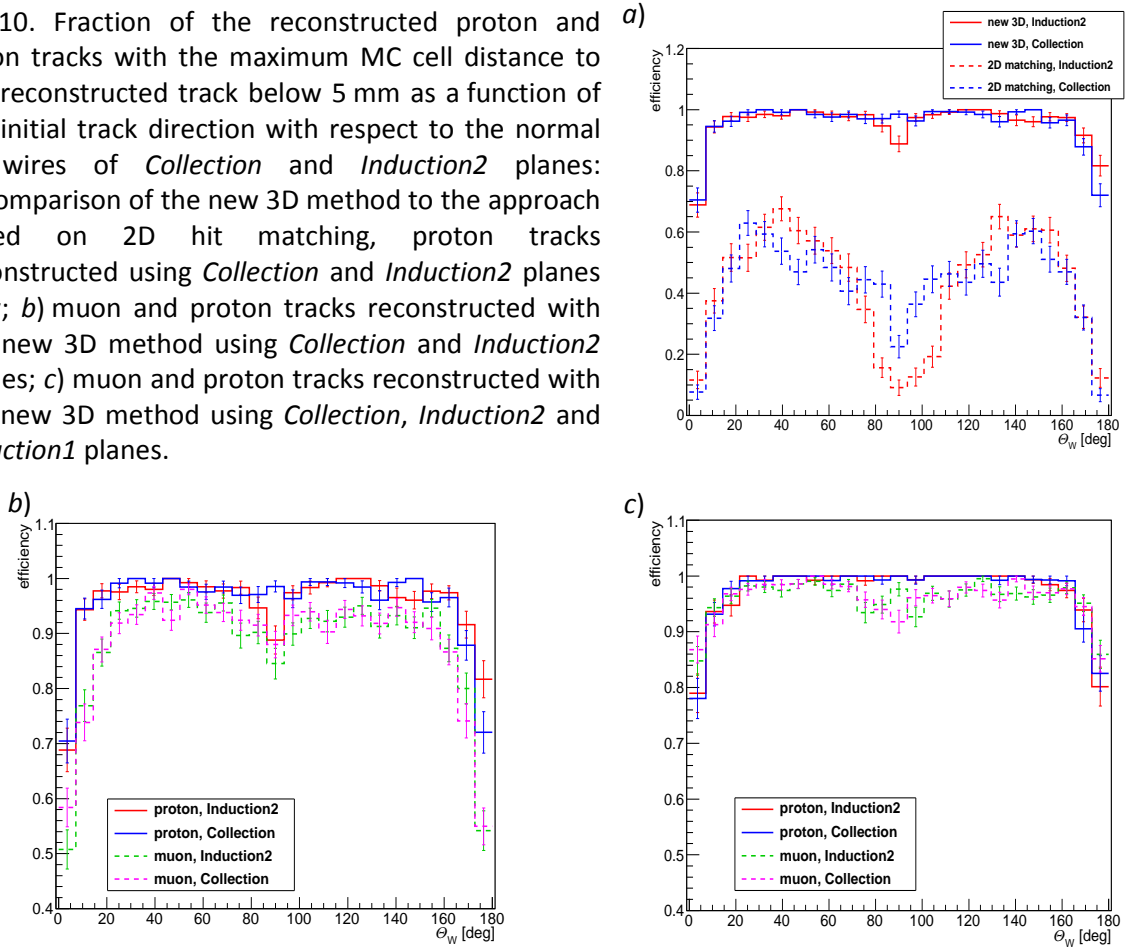

Fig. 10. Fraction of the reconstructed proton and muon tracks with the maximum MC cell distance to the reconstructed track below 5 mm as a function of the initial track direction with respect to the normal to wires of *Collection* and *Induction2* planes: *a*) comparison of the new 3D method to the approach based on 2D hit matching, proton tracks reconstructed using *Collection* and *Induction2* planes only; *b*) muon and proton tracks reconstructed with the new 3D method using *Collection* and *Induction2* planes; *c*) muon and proton tracks reconstructed with the new 3D method using *Collection*, *Induction2* and *Induction1* planes.

Adding information from the *Induction1* plane (Fig. 10c) reduces significantly spurious reconstruction, especially for the tracks parallel to the wire planes. The efficiency gain is higher in case of the muon tracks which are scattered more than protons at energies considered here and they are more likely to have track bends visible only in the XZ view. When three wire planes are used, the reconstruction fails mostly in case of straight tracks oriented strictly in the Z direction, which is more likely for proton tracks[7]. The other source of inefficiency, which is the ambiguity related to the fit initial orientation, may be solved by the correct association of the track end points, e. g. reconstructed 3D vertex associated to the track or an observation of the increasing ionization in case of particles that are stopping or decaying at rest.

3.2    Reconstruction of the particle initial direction.

Reconstruction of the particle initial direction was performed on the sample of tracks with a fixed initial kinetic energy $E_k$ (protons: 232 MeV; muons: 100 MeV) and on a sample of tracks with a range of particle initial $E_k$ (protons: 10-350 MeV; muons: 10-185 MeV). Tracks were reconstructed using the *Collection* and the *Induction2* planes.

---

[7] Also these cases may be recovered in the further development of the algorithm. A short track projection (i.e. 1-2 hits) in one of the views in the present form of the algorithm has negligible impact on the overall objective function G(F), while it should help to exclude impossible solutions. This may be achieved by a fuzzy assignment of all 2D hits to all fit segments, weighted with the distance of the hit to the segment projection in 2D.

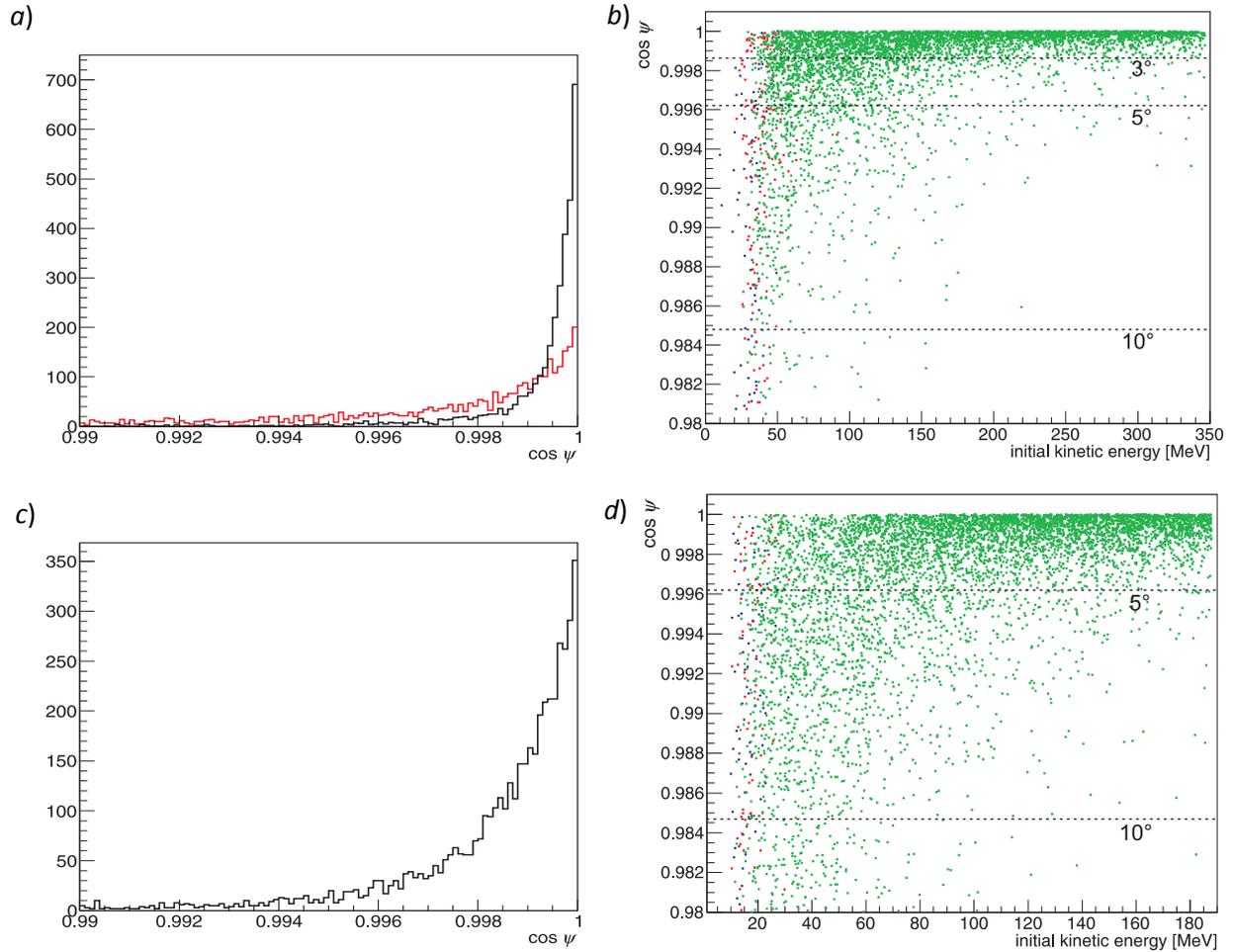

Fig. 11. Distribution of the angle $\psi$ between the reconstructed and the simulated particle initial direction: *a*) protons, initial $E_k$ = 232 MeV (range 30 cm); the new 3D reconstruction – black curve, 2D hit matching – red curve; *b*) protons, initial $E_k$: (10-350) MeV (range up to 60 cm); dots color: 1,2 hits in the track – blue, 3 hits in the track – red, 4 and more hits in the track – green; *c*) muons, initial $E_k$ = 100 MeV (range 30 cm); *d*) muons, initial $E_k$: (10-185) MeV (range up to 60 cm); colors are the same as in point *b*.

The angle $\psi$ between the reconstructed and the simulated initial directions is shown in Fig. 11. The minimum reasonable energy for proton track identification and reconstruction is about 50 MeV, which corresponds to tracks with more than 3 hits. The initial direction reconstruction at this energy threshold is better than 10° in 98% of tracks in the simulated sample. The angle $\psi$ quickly decreases for protons with higher energies and for initial $E_k$ > 232 MeV (30 cm and longer tracks) $\psi$ stays below 3° in 92% of all events. In Fig. 11a the new reconstruction method is compared with the hit matching based approach on the sample of 30 cm proton tracks. In this sample the initial direction is reconstructed with $\psi$ < 3° in 83% and 32% of events for the new method and the hit matching based approach respectively. The fraction of events with $\psi$ < 5° is 94% and 52% for the new method and the hit matching based approach respectively.

The estimation of the initial direction of muon tracks (Fig. 11c, d) is less precise due to the higher scattering along the track and lower amplitude of muon hits. The initial direction is reconstructed in the muon sample with $\psi$ < 10° for 96% of events with initial $E_k$ > 50 MeV.

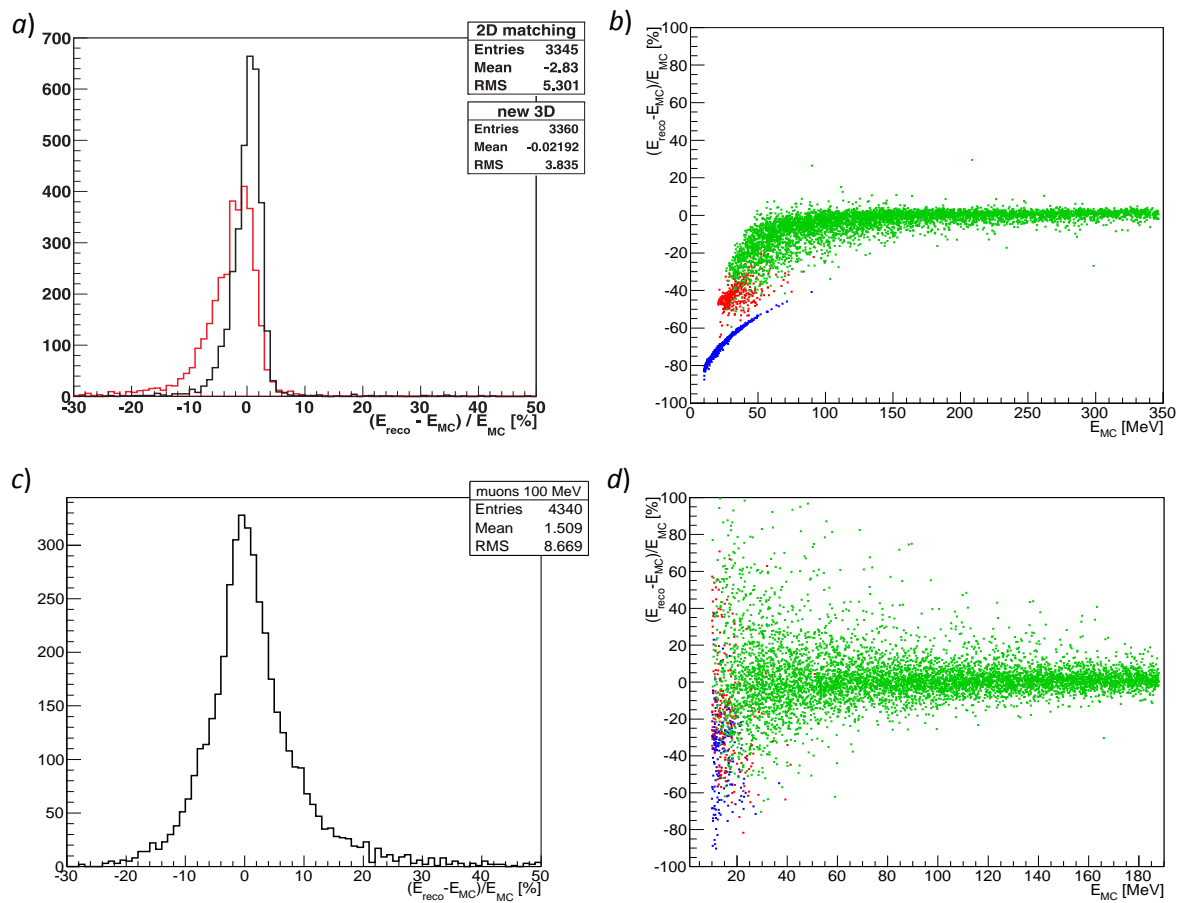

Fig. 12. Calorimetric reconstruction: *a)* protons, initial $E_k$ = 232 MeV, range 30 cm; the new 3D reconstruction – black curve, 2D hit matching – red curve; *b)* protons, initial $E_k$: (10-350) MeV, range up to 60 cm; dots color: 1,2 hits in the track – blue, 3 hits in the track – red, 4 and more hits in the track – green; *c)* muons, initial $E_k$ = 100 MeV, range 30 cm; *d)* muons, initial $E_k$: (10-185) MeV, range up to 60 cm; color the same as in point *b*.

For events with the initial $E_k$ > 100 MeV (30 cm and longer tracks) the initial direction is reconstructed with $\psi$ < 3° in 67% of events and with $\psi$ < 5° in 90% of events.

3.3   Calorimetric reconstruction.

The calorimetric reconstruction was performed on the track samples used in the previous tests. The hit charges along the track were corrected for the recombination effect according to the procedure given in Section 2.2. However, the charges of the first hit and the last hit in the track need a different approach due to the high uncertainty of the outermost d$x$ lengths, where the particle partially crosses the distance between consecutive wires. These charges were corrected in a conservative way, using a constant factor corresponding to the minimum ionizing particle, R = 0.7, according to Eq. 8. The results of the particle energy reconstruction are shown in Fig. 12.

In the presented test we assume that the particle is not identified otherwise it would allow for the differentiation of the correction factor R applied to the outermost hits. Therefore the treatment of these hits has a strong effect in case of the highly ionising, low energy protons whose tracks are composed of a few hits. In such cases the underestimated charge of the outermost hits has a major contribution to the overall particle energy calculation, as seen in Fig. 12b. The effect is decreasing with the growing length of the track, however it is still visible as the asymmetry of the distribution of the reconstruction error of the proton tracks with initial $E_k$ = 232 MeV (corresponding to 30 cm tracks), shown in Fig. 12a. The energy underestimation is observed also in case of low energy muon tracks, but in a much smaller extent (Fig. 12d).

The energy reconstruction of the tracks with low amplitude hits, present for minimum ionizing muons, is limited by the electronics noise that affects both the hit charge and the hit position reconstruction. This leads to larger errors of the recombination correction and of the energy reconstruction of the muon tracks than in case of proton tracks. The asymmetry in the energy reconstruction error distribution of the muon tracks in Fig. 12c is originating from the nonlinearity of the recombination correction.

Results of the energy reconstruction are compared for tracks obtained from the new method with those obtained from the hit matching approach, using the sample of proton tracks with initial $E_k$ = 232 MeV (Fig. 12a). More accurate reconstruction of the particle trajectory allows application of more precisely calculated recombination correction.

**4. Tests on data tracks.**

The new reconstruction procedure was applied to tracks of stopping particles selected from data collected with the CNGS beam. The correction of the recombination effect was applied according to the procedure described in Section 2.2. The goal of the test was to compare the theoretical Bethe-Bloch curves describing $dE/dx$ evolution along the stopping particle track with the $dE/dx$ sequence reconstructed for data tracks. At the same time the performed analysis is a test of the Birk's low application to the energy correction due to the recombination effect [12].

The main attention had been paid to low energy protons because they are relatively easy to be recognized in manual scanning: they are highly ionizing, not decaying particles. 295 tracks in total were selected visually, according to the following criteria:
1. there are no visible decay products;
2. ionization is increasing at the track end;
3. hits in the *Collection* view are not overlapping with other tracks and cascades;
4. track has at least 5 hits in the *Collection* view.

The above conditions select mostly the stopping proton tracks, with a small fraction of the protons interacting inelastically at low energies producing neutral or undetectable charged secondaries. The sample contains also a fraction of muons and pions absorbed at rest. Kaons are unlikely to meet the selection criteria. The fraction of kaon tracks that could be misidentified as stopping particles, with no visible secondary particles, is about 0.3% according to the FLUKA simulation with the selection criteria applied as for data tracks.

Tracks from the selected sample were reconstructed automatically and the sequence of $dE/dx$ values was evaluated. The particle identification is based on the $dE/dx$ versus track range, which is computed from the point where the particle stops, called later residual range. The particle identification was performed in order to distinguish tracks compatible with the proton hypothesis from the μ/π hypothesis. A detailed description of the particle identification procedure will be given in a forthcoming paper. The results in comparison with the theoretical curves are shown in Fig. 13a. The theoretical stopping power curves of Bethe-Bloch have been calculated taking into account LAr's properties, shell corrections and the density effect. The delta rays produced by particles in the examined momenta range have an energy low enough to be undistinguishable from the particle track, therefore no delta rays energy restriction has been set. Tracks compatible with μ/π hypothesis are well distinguishable from the main band of the proton contribution, as it is shown in Fig. 13a. Fig. 13b-d present comparison of the reconstruction of the data tracks and the simulated proton and muon tracks. The small shift of the proton data d*E*/d*x* distribution toward the lower values may originate from the interacting protons contamination in the selected track sample (Fig. 13c).

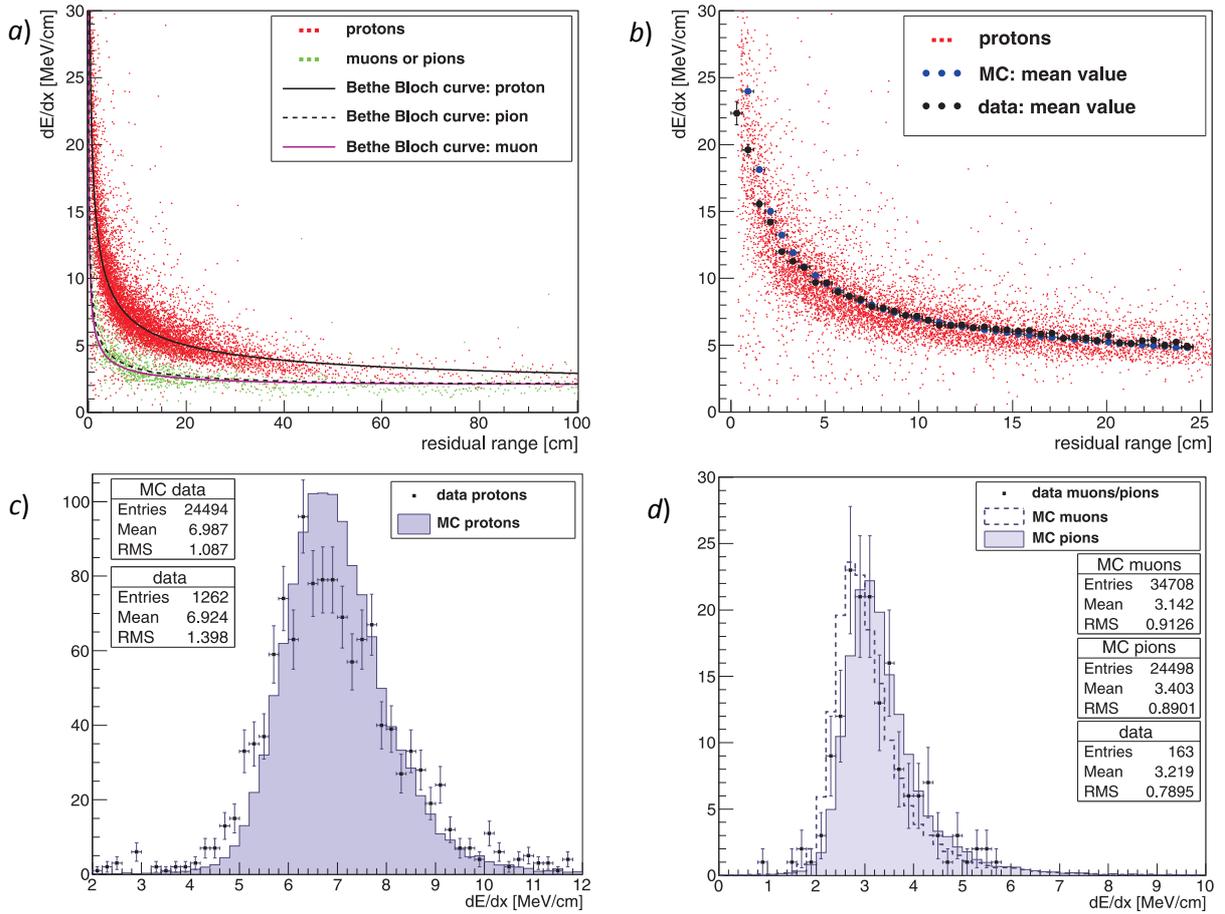

Fig. 13. *a*) The selected sample of data tracks identified as proton or μ/π tracks; *b*) mean value of d*E*/d*x* distributions in 6 mm bins of *residual range*: black – data tracks identified as protons; blue – simulated proton tracks; points in the background – data tracks identified as protons; *c*) d*E*/d*x* in 8-12 cm *residual range* for tracks identified as protons, compared to simulated proton tracks; *d*) d*E*/d*x* in 8-12 cm *residual range* for tracks identified as μ/π, compared to simulated muon and pion tracks.

In addition to the described data analysis, we show the example of a kaon decay which was identified in one of the CNGS events and it is shown in Fig. 14a, b, in *Collection* view and *Induction2* view respectively. This topology is characteristic for the proton decay searches in the channel $p \rightarrow K^+ \bar{\nu}$. Hits that belong to the chain of the decay (kaon, muon and electron/positron) were used in the reconstruction as they were a single track. The 3D view of the obtained tracks is shown in Fig 14c. Even though in the *Induction2* the decay of kaon into muon happened at the narrow angle and the electron/positron is crossing the kaon track, the 3D topology is resolved.

Since the kaon and the muon decay at rest, it is possible to identify them according to the dependence of d*E*/d*x* versus residual range along the tracks. The particle identification description is out of the scope of this paper and here we present only the measurement of d*E*/d*x* sequence of both the kaon and the muon track, Fig. 14d. The measurement was done selecting manually the end points of tracks. In case of the muon, 30 cm of the track residual range was used to distinguish it between other particles. In case of kaon, the track residual range was taken from the point of the particle decay to the point of the elastic scattering, which is seen as a kinked trajectory on Fig. 14a-c, and which corresponds to 8.5 cm of the track residual range.

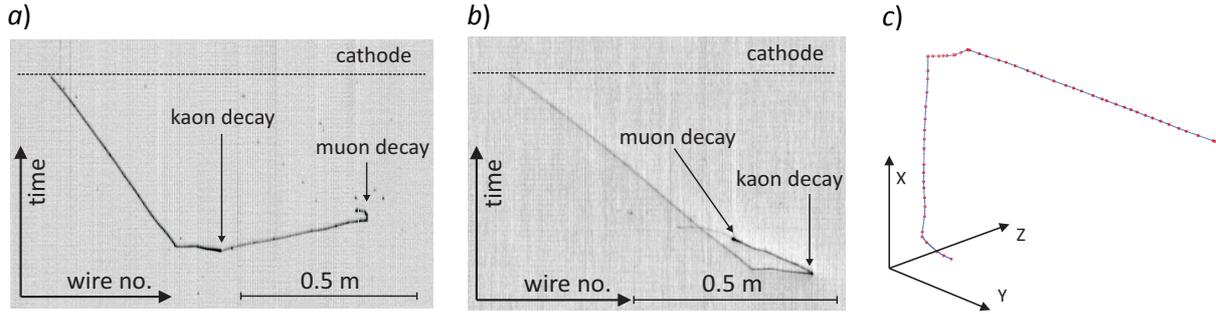

Fig. 14. Example of a decaying kaon candidate observed in the CNGS data (K: 90 cm, 325 MeV; μ: 54 cm, 147 MeV; e: 13 cm, 27 MeV): *a)* the *Collection* view; *b)* the *Induction2* view; *c)* the 3D reconstruction (fit nodes are marked with the red dots); *d)* d$E$/d$x$ sequence for the kaon track (cyan dots) and the muon track (violet dots) superimposed on the theoretical Bethe-Bloch curves. The muon data points with the high d$E$/d$x$ are due to the additional energy of delta rays. They are attached to the muon track (visible black dots on the muon track in the *Collection* view). Their energy contribution is in agreement with the energy loss distribution expected for a muon track.

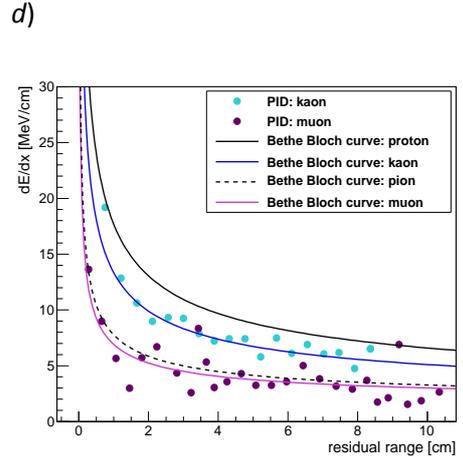

## 5. Summary and future development remarks.

A new approach to the 3D reconstruction for LAr TPC detectors was proposed and applied to the reconstruction of data tracks collected with the ICARUS T600 detector. The main advantage of the proposed idea is the full exploitation of all available pieces of information. The reconstructed object is built in the 3D space to match data simultaneously in all its 2D projections, with a set of object-specific constraints, e. g. the curvature and the distance to 3D vertices in the track reconstruction. In contrast to the usual approach in former works related to LAr TPC, the troublesome matching of 2D points by the drift timing is no longer needed. Moreover, in case of the application to the track reconstruction:
  a) 2D hits projected to the reconstructed 3D trajectory are not constrained to the wire crossing points in the XZ plane and allow precise hit-by-hit analysis of the charge deposition;
  b) missing parts of the track in one of the wire planes are properly taken into account;
  c) the method is much more efficient than hit matching approach in the reconstruction of tracks parallel to wire planes.

The efficiency of the proposed track reconstruction method was demonstrated on samples of simulated stopping proton and muon tracks. We also presented the results of the application to data, such as d$E$/d$x$ reconstruction on a sample of stopping particles selected from CNGS neutrino interactions observed with ICARUS T600 detector. The presented data tests show a very good agreement with the MC simulation and the theoretical expectations of the $dE/dx$ evolution of the stopping particles.

In this paper we focused on the track reconstruction, however the projection, the distance and the constraint operators (Eq. 1) may be constructed for showers, vertices or any other object. For example, an electromagnetic shower axis may be reconstructed as a single 3D segment optimized to the hits in wire projections, similarly to the initialization procedure given in Section 2.2.1. Another example could be the simultaneous optimization of several tracks with the common starting fit node which acts as the interaction vertex. These ideas are now under study. Issues specific to the detector readout design such as merging objects from multiple TPC modules or LEM units may be accommodated by the algorithm as well. We consider the proposed approach as the optimal utilization of all the spatial information contained in the LAr TPC data.

**Acknowledgments.**

The ICARUS Collaboration acknowledges the fundamental contribution of INFN to the construction and operation of the experiment. In particular we are indebted to the LNGS Laboratory for the continuous support to the experiment. The Polish groups acknowledge the support of the Ministry of Science and Higher Education in Poland, including project 637/MOB/2011/0. Finally we thank CERN, in particular the CNGS staff, for the successful operation of the neutrino beam facility.

**References.**

1. C. Rubbia, *The Liquid-Argon Time Projection Chamber: a new concept for neutrino detectors*, CERN Report (1977)
2. C. Anderson et. al., *The ArgoNeuT Detector in the NuMI Low-Energy beam line at Fermilab*, JINST 7 (2012) P10019
3. J-PARC T32 Collaboration, *Performance of a 250L liquid Argon TPC for sub-GeV charged particle identification*, arXiv 1206.1181v1 (2012)
4. MicroBooNE Collaboration, *Proposal for a New Experiment Using the Booster and NuMI Neutrino Beamlines: MicroBooNE*, FERMILAB-PROPOSAL-0974 (2007)
5. D. Autiero et. al., *Large underground, liquid based detectors for astro-particle physics in Europe: scientific case and prospects*, JCAP11 (2007) 011
6. ICARUS Collaboration, *Underground operation of the ICARUS T600 LAr-TPC: first results*, JINST 6 (2011) P07011
7. ICARUS Collaboration, *Design, construction and tests of the ICARUS T600 detector*, Nuclear Instruments and Methods in Physics Research A527 (2004) 329
8. A. Badertscher et. al., *First operation of a double phase LAr Large Electron Multiplier Time Projection Chamber with a two-dimensional projective readout anode*, arXiv 1012.0483 (2010)
9. B. Kegl, A. Krzyzak, T. Linder, K. Zeger, *Learning and design of principal curves*, IEEE Transactions on Pattern Analysis and Machine Intelligence Vol. 22 3 (2000) 281-297
10. ICARUS Collaboration, *Analysis of the liquid argon purity in the ICARUS T600 TPC*, Nucl. Inst. Meth., A516 (2004) 68-79

11. M. Miyajima et al., *Average energy expended per ion pair in liquid argon*, Phys. Rev. A9 (1974) 1438
12. ICARUS Collaboration, *Study of electron recombination in liquid argon with the ICARUS TPC*, Nuclear Instruments and Methods in Physics Research A523 (2004) 275
13. G. Battistoni, S. Muraro, P.R. Sala, F. Cerutti, A. Ferrari, S. Roesler, A. Fasso, J. Ranft, Proceedings of the Hadronic Shower Simulation Workshop 2006, Fermilab 6-8 September 2006, M. Albrow, R. Raja eds. AIP Conference Proceeding 896 (2007) 31-49
14. A. Ferrari, P.R. Sala, A. Fasso, J. Ranft, *FLUKA: a multi-particle transport code*, CERN-2005-10 (2005) INFN/TC_05/11, SLAC-R-773